\documentstyle[10pt,emulateapj]{article}        
\def\deg{$^{\circ}\,$}
\def\solm{M$_{\odot}\,$}

\begin{document}     
\vspace{-7cm}
\title{Evidence for a Major Merger Origin of High Redshift Sub-millimeter Galaxies}

\author{Christopher J. Conselice$^{1}$, Scott C. Chapman$^{2}$, Rogier A.
Windhorst$^{3}$}

\altaffiltext{1}{California Institute of Technology, MS 105-24, Pasadena
CA 91125}
\altaffiltext{1}{NSF Astronomy \& Astrophysics Postdoctoral Fellow}
\altaffiltext{2}{California Institute of Technology, MS 320-47, Pasadena, CA 91125}
\altaffiltext{3}{Arizona State University, Dept. of Physics and Astronomy, Tempe, AZ, 85287-1504}

\begin{abstract}

Sub-mm detected galaxies located at redshifts $z > 1$ host a major
fraction of the bolometric luminosity at high redshifts due to
thermal emission from heated dust grains, yet the nature of these
objects remains a mystery. The major problem in understanding 
their origin is whether the dust heating mechanism is predominantly caused
by star formation or active galactic nuclei, and what triggered this 
activity.    
We address this issue by examining the structures of a sample of
11 sub-mm galaxies imaged with STIS on the Hubble Space Telescope. We 
argue that $\sim 61\pm21$\% of these sub-mm sources are undergoing an 
active major merger using the CAS quantitative morphological system.  We
rule out at $\sim$ 5 $\sigma$ confidence that these sub-mm galaxies
are normal Hubble types at high redshift.  This merger fraction
appears to be higher than for Lyman-break galaxies 
undergoing mergers at similar redshifts. 
Using reasonable constraints on the stellar masses of Lyman-Break
galaxies and these sub-mm sources, we further argue that at redshifts
$z \sim 2-3$ systems with high stellar masses are more likely than lower
mass galaxies to be involved in major mergers.

\newpage

\end{abstract}

\section{Introduction}

Faint sub-mm galaxies were discovered with the first generation of 
multi-element detecting devices working at sub-mm wavelengths, most
notably the SCUBA array (Smail, Ivison \& Blain 1997).  The nature of these
galaxies has however remained a mystery, despite a considerable amount of
observing time spent obtaining $\sim 300$ detections in the field
and behind lensing clusters (Blain et al. 2002).

The generally accepted working idea is that these sub-mm galaxies are
distant $z > 1$ systems which emit in the rest-frame far infrared
due to thermal emission from dust grains heated by star formation and/or
active galactic nuclei (AGN).  These objects are also thought to be analogs 
of nearby ultra luminous
infrared galaxies (ULIRGs), although even in nearby ULIRGs 
it is debated whether or not the
dust is heated by the UV continuum of massive
stars, or an active nucleus.    As sub-mm
galaxies are 400 times more common at $z \sim 2$ than at
$z \sim 0$, they constitute a significant fraction of the bolometric 
luminosity density at high redshift (Chapman et al. 2003a), thus
understanding their origin and relationship to nearby galaxies is of
central importance.
 
Answering fundamental questions concerning sub-mm galaxies by studying them 
at multiple wavelengths 
has remained very difficult due to the low-resolution of SCUBA and
other sub-mm instruments, making identifications at other wavelengths
difficult.  One approach to this problem has been to identify
sub-mm galaxies through their emission in the radio (Ivison et al. 1998; 
Chapman et al. 2002a), utilizing the fact
that sub-mm/far-infrared luminosities correlate in nearby galaxies 
(Helou et al. 1985).  Through these identifications, we are able to
determine sub-mm source positions to within a sub-arcsec, from
which optical follow-up can be done readily.  A subset of these
sub-mm galaxies whose positions were located using the radio have
been imaged with the Hubble Space Telescope using STIS (see also a companion
paper, Chapman et al. 2003b).  Previous studies have been largely qualitative,
and often with misidentifications (e.g., Smail et al. 1998; Ivison et al. 
2001; Chapman et al. 2002b).  

Based on our analysis we conclude that 40 - 80\% of these
sub-mm sources are consistent with undergoing a major merger, and statistically
rule out at 5 $\sigma$ confidence that these systems are normal galaxies at
 high redshift.  We
further demonstrate, through comparisons with artificially redshifted
nearby ULIRGs and normal (non-ULIRG) galaxies, that high redshift sub-mm 
galaxies appear qualitatively similar to ULIRGs, and are potentially forming
into massive spheroids.

\section{Imaging and Analysis Method}

Our sample consists of 11 sub-mm sources selected in the radio.   The
full sample selection for these galaxies is described in 
Chapman et al. (2003b).  Although there are some selection issues based on
using radio identifications, these objects were not selected based on 
optical properties, and they span
a wide range of optical magnitudes (Chapman et al. 2003b).
Each of the sub-mm sources (Figure~1) we study was imaging in the 
50CD clear filter in one to three orbits, in two separate exposures per 
orbit, resulting in a total exposure time of 1-3 ks
(see Chapman et al. 2003b for further details).   

We analyzed these images using the CAS (concentration, asymmetry and
clumpiness) morphological system (Conselice 2003).  In the CAS
system, the asymmetry index ($A$) is used to determine whether or not
a galaxy is involved in a major merger (Conselice et al. 2000a,b; Conselice
2003; Conselice et al. 2003).  The value of $A$ is calculated by rotating
a galaxy through 180\deg and subtracting this rotated galaxy from the original
and comparing the absolute value of the residuals of this subtraction to
the original galaxy flux (Conselice et al. 2000a).
Before carrying out the CAS analysis on the sub-mm galaxy
images we removed any galaxies and stars
close to the sub-mm sources, so as to not contaminate the measurements.
We placed our initial guess for the center on the brightest portion of
the sub-mm galaxy, and then ran the CAS program to determine 
asymmetries and light concentrations.

To understand the systematics due to a lowered resolution and signal to
noise we artificially redshifted 50 nearby 
ULIRGs (from Farrah et al. 2001 and Conselice 2003) and 82 normal Hubble 
types (Frei et al. 1996, Conselice 2003) to how they would appear in our STIS 
images at $z \sim 2-3$, the spectroscopic redshift range of these sources 
(Chapman et al. in prep).  These simulations are done by
reducing the resolution and surface brightness of a galaxy as to how it
would appear at high redshifts.  The amount of noise matching that expected 
in the STIS observations are then added in.   

Once we create these new 
images, based on these simulations, we remeasure the CAS parameters of the
simulated galaxies, in the same manner as on the original sub-mm sources.   
  Note that we are taking a
very conservative approach by including these two types of nearby
galaxies to account for redshift effects.  The apparent morphology of the
sub-mm sources is very peculiar and therefore the likewise peculiar 
nearby ULIRGs are a better population for making this correction.  Normal 
galaxies also do not become significantly more irregular
in the rest-frame ultraviolet (Windhorst et al. 2002).  We
use the nearby normal sample as an extreme lower limit to what these 
corrections could be. 

\section{Results}

\subsection{Evidence For a Merger Origin}

Concentration-asymmetry diagrams for our sample of sub-mm galaxies
are shown in Figure~2.   The asymmetry and concentration values
for these galaxies have been corrected for redshift effects assuming they 
have morphologies intrinsically similar to ULIRGs (Figure~2a,c), and nearby 
normal galaxies (Figure~2b,d) (Conselice et al. 2000a).   

We also plot on Figure~2a,b the average values 
and $1 \sigma$ variations of measured concentrations and asymmetries for
nearby galaxy populations, including ULIRGs, as observed
in the rest-frame optical (Conselice 2003). Although we view these sub-mm
galaxies in the near to mid-UV, the morphological
appearance of ULIRGs and other star forming galaxies does not
significantly differ between optical and near-UV wavelengths 
(e.g., Conselice et al. 2000c; Surace \& Sanders 2000; Windhorst et al. 2002).
Figure~2c,d shows the location of the sub-mm galaxies in reference to the 
actual
$C-A$ values found for nearby normal galaxies, including separately labeled
ellipticals and ULIRGs. 

Figure~2 can be used to determine which population the sub-mm 
galaxies are most similar to morphologically. As can be seen in Figure~2,
the dominate morphological 
feature of the sub-mm galaxies is their high asymmetries. The distribution of
the sub-mm sources in $C-A$ space is also most similar to the ULIRGs, and does
not overlap much with any normal galaxy type.  Using the
major merger criteria calibrated in Conselice (2003), based on
nearby ULIRGs, we find that a galaxy is likely a major merger if 
$A > A_{\rm merger} = 0.35$.
Using the ULIRG z $\sim$ 2 correction, we find that the merger
fraction for the sub-mm sources is 0.82, which is the same fraction found
when correcting by the $z \sim 3$ simulations.
The normal galaxy correction still reveals a large merger fraction of
$\sim 0.4$, which is certainly a lower limit to the actual fraction of
galaxies involved in mergers.  This lower limit is as high as the largest 
merger fractions found for any Lyman-Break galaxy population 
(Conselice et al. 2003) (\S 3.2).   

There are naturally uncertainties in this merger fraction calculation. First,
the sample size is small, only 11 galaxies.  Second, it is based on a
correction which has necessarily some uncertainty in it, since the sub-mm
galaxies may be a mix of normal and merger/ULIRG systems.  To
circumvent this problem we performed a series of Monte Carlo simulations
using our normal and ULIRG galaxy samples after they were simulated
to $z \sim 2$.  We take at random 11 of these objects and compute their 
asymmetry distributions, and compare this to the asymmetry distributions of
the sub-mm sources.  Doing this, we find at a 5~$\sigma$ confidence that
normal galaxies at $z \sim 2$ cannot reproduce the asymmetry distribution of
the sub-mm sources.  The simulated ULIRGs are also slightly less asymmetric
than the sub-mm sources, but have asymmetry values distributions
consistent to within 1.9 $\sigma$.

\subsection{Sub-mm and Lyman-break Galaxies}

The implied merger fraction of these sub-mm galaxies appears to be
slightly larger than what is 
found for UV bright galaxies in the Hubble Deep Field (HDF)
(Conselice et al. 2003) at similar redshifts.
Lyman-break galaxies at $z \sim 2.5$ found in the HDF
all have implied merger fractions lower than our derived sub-mm values
(Conselice et al. 2003).  For example UV bright galaxies at $2 < z < 3$
with magnitudes M$_{\rm B} < -20$ and stellar masses $> 10^{9.5}$ \solm
have a merger fraction $\sim 0.18$ (Figure~3). The brightest and most 
massive galaxies seen in the HDF, with M$_{\rm B} < -21$ and
M$_{*} > 10^{10}$ \solm, have merger fractions $\sim 0.4 - 0.5$, which
is lower than the implied merger fractions of our sub-mm sample.
From this it appears that the sub-mm galaxies are actively undergoing
mergers in a greater abundance than Lyman-break galaxies. 
This may be the result of higher mass systems 
undergoing more mergers at higher redshifts (Figure~3).   Sub-mm galaxies as a 
population appear to
be dominated by systems involved in major mergers, while the Lyman-break 
galaxies are more likely to be in various phases of evolution,
and perhaps includes starbursts not induced by major mergers.  On Figure~3 we
plot the merger fraction of the sub-mm sources at a stellar
mass of 10$^{11}$ \solm.  This estimate is likely roughly correct as
sub-mm galaxies have dynamical and gas masses $> 10^{11}$ \solm (e.g.,
Frayer et al. 1998).

\section{Comparison to z$\sim 0$ Galaxies: ULIRGs and Spheroids}

Although we are in the regime of small number statistics, we can argue that
the sub-mm galaxies are similar to nearby ULIRGs in terms of their structural 
properties, in addition to their already well established similarities in
producing rest-frame far infrared light (e.g., Dey et al. 1999).

We have already argued this through the high asymmetries and
implied high merger fraction for the sub-mm galaxies, and the low
probability that these sub-mm sources are morphological
similar to normal galaxies.  This can be shown qualitatively as
well.  Figure~4 shows
nearby ULIRGs at $z \sim 0.1$, and the same galaxy after it has been
simulated to how it would appear at $z \sim 3$.  These are the same
simulations discussed in \S 2 to determine the correction to the
concentration and asymmetry values for the sub-mm galaxies.  These
simulated ULIRGs appear very similar to the
sub-mm galaxies themselves (Figure~1), although the separation
between components in nearby ULIRGs and the sub-mm sources
differ (Chapman et al. 2003b).  Nearby normal galaxies, such
as disks and ellipticals do not appear in a similar manner when redshifted
out to these same redshifts.

We can further test how similar the ULIRGs and sub-mm galaxies are
in terms of their morphologies by performing a K-S test on their
asymmetries and concentrations.   The probability that the sub-mm
source asymmetries are taken from the ULIRG population asymmetries are
21\% after correcting by the ULIRG simulation results, and 2\% when 
correcting by the 
normal galaxy simulations.  The probability of association with the
normal galaxies is $< 0.001$\% when using the ULIRG correction and
1\% when using the normal galaxy simulation correction.  The concentration
index probabilities of association with ULIRGs are 96\% for the ULIRG 
correction and 78\% for the normal galaxy correction.    The corresponding
probabilities for a normal galaxy association are 12\% and 32\%, respectively.
Although these K-S tests are not conclusive they again rule out that
the sub-mm asymmetry distribution is similar to the asymmetry
distribution of normal galaxies.  

If sub-mm sources are analogs of nearby ULIRGs then they are perhaps
forming into modern ellipticals.  Ellipticals can be uniquely identified
in the CAS system by their high light concentrations and low
asymmetries.  The only galaxies that have light concentrations
as high as ellipticals are ULIRGs (Bershady et al. 2000; Conselice 2003), 
suggesting a causal connection.
Based on Figure~2 it can also be seen that some of the sub-mm
sources have light concentrations as high as ellipticals, suggesting
that at least some of these systems are forming into
spheroids.   We can conclusively rule out that all of these systems are 
spirals, as 4 (36\%) have light concentrations too high to 
be forming into disks.

\section{Discussion and Conclusions}

Sub-mm galaxies and ULIRGs have comparable
amounts of rest-frame far infrared luminosities, and large masses,
suggesting that sub-mm sources result from a similar merger origin.  
However, direct quantitative
evidence has been lacking.  We argue in this paper that a significant
fraction of sub-mm sources
are galaxies actively engaged in major mergers by using the
CAS morphological system (Conselice 2003).  On-going major mergers are defined
as systems with an asymmetry higher than the locally calibrated limit of
$A_{\rm merger} = 0.35$.  Using this limit, and
correcting for redshift effects, we find that 40-80\% of sub-mm galaxies
are consistent with undergoing major mergers.  These results have important 
implications for the nature of massive galaxy formation.  Sub-mm sources 
constitute a significant
galaxy population at high redshift, undergoing mergers, 
that likely later evolve into the massive galaxies seen in the nearby universe.
These systems are also involved in large amounts of star formation that
may be induced by the major merger, demonstrating that at least for the most
massive galaxies, their stars are formed through merger induced starbursts
rather than collapses.


\begin{figure}
\plotfiddle{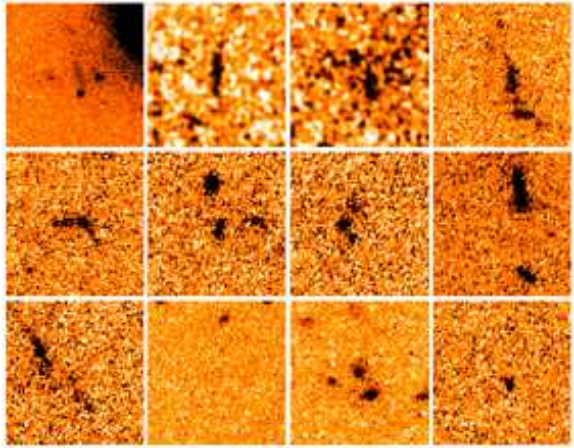}{6.0in}{0}{80}{80}{-250}{-100}
\vspace{0in}
\caption{Montage images of the 12 sub-mm galaxies imaged with STIS on
the Hubble Space Telescope.  In this paper, we examine the quantitative 
morphological properties of all but one of these systems whose structure
is too faint for an analysis. }
\end{figure}

\begin{figure}
\plotfiddle{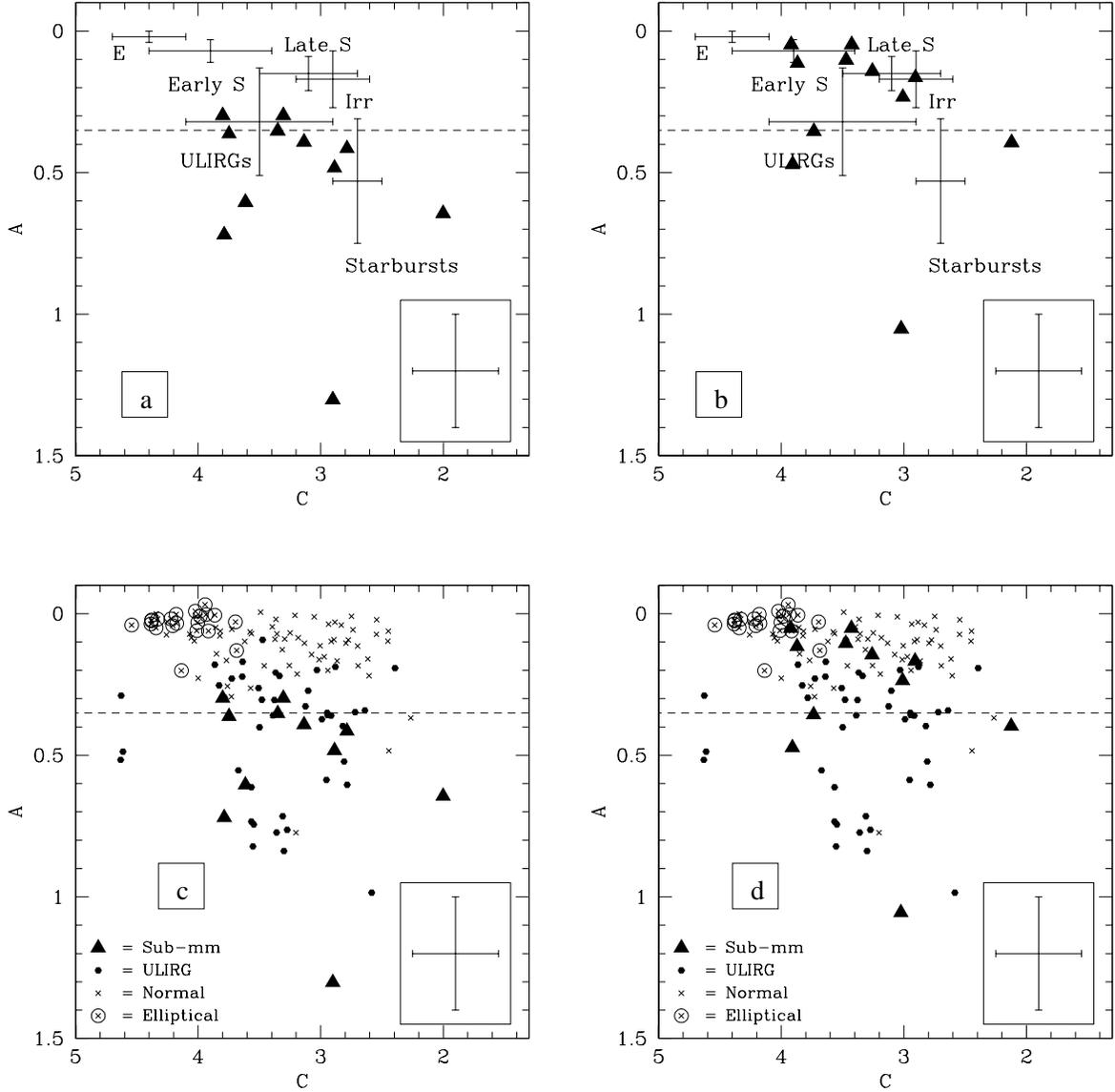}{6.0in}{0}{80}{80}{-250}{-100}
\caption{Concentration ($C$)-asymmetry ($A$) diagram for the sub-mm galaxies
studied in this paper.  The triangles show the locations of the sub-mm
sources on these diagrams after (a,c) correcting their measured values using
$z \sim 0$ ULIRGs simulated to $z = 2$ and (b,d) correcting values
using $z \sim 0$ normal galaxies simulated to $z \sim 2$.  
The various crosses in (a,b) show where average nearby galaxy populations 
fall in this space and their 1 $\sigma$ variations. Plotted in (c,d) are
the individual $C-A$ values for
nearby ULIRGs, and normal galaxies, with the ellipticals circles.  The 
horizontal dashed line is the asymmetry
limit for major mergers, such that a galaxy is likely undergoing a merger if
$A > A_{\rm merger}$ = 0.35. }
\end{figure}

\begin{figure}
\plotfiddle{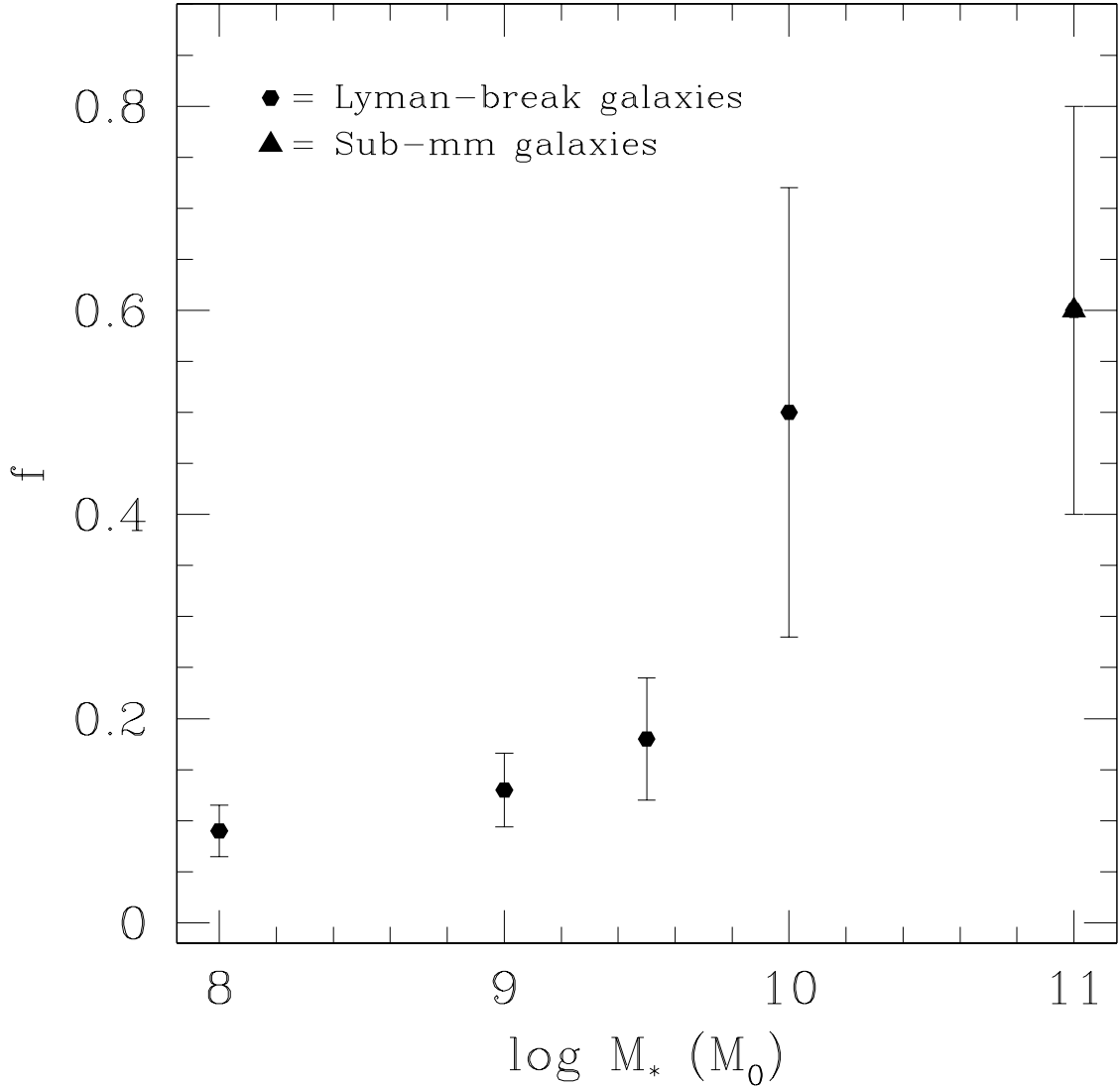}{6.0in}{0}{80}{80}{-250}{-100}
\caption{Fraction of galaxies between $2 < z < 4$ consistent with undergoing a 
major merger, (f), as a function of stellar mass lower limit.  The stellar 
masses of the 
Lyman-break galaxies (LBGs) were obtained through fits to spectral
energy distributions acquired by Papovich et al. (2001) and whose merger
properties are discussed in detail in Conselice et al. (2003).  The stellar 
mass lower limit of 10$^{11}$ \solm for the sub-mm galaxies is estimated 
based on the large gas and dynamical mass measurements of these galaxies.}
\end{figure}

\begin{figure}
\plotfiddle{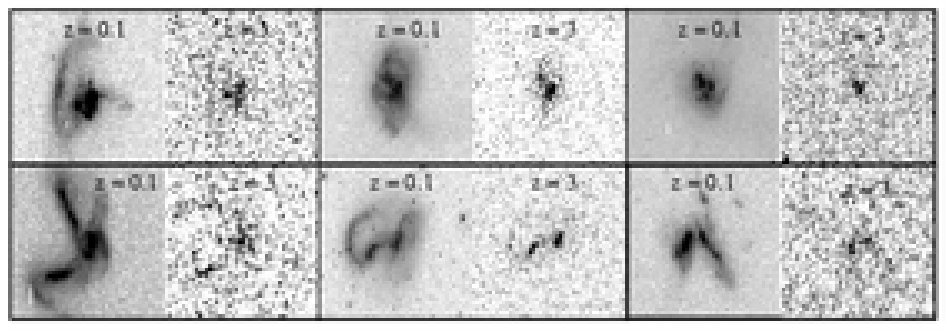}{6.0in}{0}{80}{80}{-250}{-100}
\vspace{-2in}
\caption{Nearby ULIRGs at $z \sim 0.1$, and images of the same galaxies
after they have been simulated to $z \sim 3$, and viewed under the same
conditions as the sub-mm galaxies.}
\end{figure}
 

\begin{references}

\reference{} Bershady, M.A., Jagren, A., \& Conselice, C.J. 2000, AJ, 119, 2645\
\reference{} Blain, A.W., Smail, I., Ivison, R.J., Kneib, J-P., \& Frayer, D.T. 2002, Physics Reports, 369, 111
\reference{} Chapman, S.C., Lewis, G.F., Scott, D., Borys, C., \& Richards, E. 2002a, ApJ, 570, 557 
\reference{} Chapman, S.C., Shapely, S., Steidel, C., \& Windhorst, R. 2002b, ApJ, 572, 1L
\reference{} Chapman, S.C., et al. 2003a, Nature, in press
\reference{} Chapman, S.C., et al. 2003b, ApJ, submitted
\reference{} Conselice, C.J., Bershady, M.A., \& Jangren, A. 2000a, ApJ, 529, 886
\reference{} Conselice, C.J., Bershady, M.A., \& Gallagher, J.S. 2000b, A\&A, 354, 21L
\reference{} Conselice, C.J., Gallagher, J.S., Calzetti, D., Homeier, N., Kinney, A. 2000c, AJ, 119, 79
\reference{} Conselice, C.J. 2003, ApJS, 147, 1
\reference{} Conselice, C.J., Bershady, M.A., Dickinson, M., \& Papovich, C. 2003, AJ, in press, astro-ph/0306106
\reference{} Dey, A., Graham, J.R., Ivison, R.J., Smail, I., Wright, G.S., \& Liu, M.C. 1999, ApJ, 519, 610
\reference{} Farrah, D., et al. 2001, MNRAS, 326, 1333
\reference{} Frayer, D., et al. 1998, ApJ, 506, L7
\reference{} Frei, Z., Guhathakurta, P., Gunn, J.E., \& Tyson, T.J. 1996, AJ, 111, 174
\reference{} Helou, G., Soifer, B.T., \& Rowan-Robinson, M. 1985, ApJ, 298, 7L
\reference{} Ivison, R.J. et al. 1998, ApJ, 494, 211
\reference{} Ivison, R., Smail, I., Frayer, D., Kneib, J.-P., \& Blain, A.W. 2001, ApJ, 561, L45
\reference{} Papovich, C., Dickinson, M., \& Ferguson, H.C. 2001, ApJ, 559, 620
\reference{} Smail, I., Ivison, R.J., \& Blain, A.W. 1997, ApJ, 490, 5L
\reference{} Surace, J., \& Sanders, D.B. 2000, AJ, 120, 604
\reference{} Windhorst, R.A. 2002, ApJS, 143, 113

\end{references}
\end{document}